\documentclass[runningheads,numberwithinsect]{llncs}

\usepackage{hyperref}
\usepackage{url}

\usepackage{xargs}
\usepackage[pdftex,dvipsnames]{xcolor} 

\usepackage[colorinlistoftodos,color=red!10,prependcaption,textsize=small
]{todonotes}

\usepackage{graphicx}

\newcommandx{\reminder}[2][1=]{}

\begin{document}

\title{Check Your (Students') Proofs---With Holes
}

\author{Dennis Renz\and
  Sibylle Schwarz\and
  Johannes Waldmann
}
\institute{HTWK Leipzig}

\maketitle

\begin{abstract}
  Cyp (Check Your Proofs) (Durner and Noschinski 2013; Traytel 2019)
  verifies proofs about Haskell-like programs.
  We extended Cyp with a pattern matcher
  for programs and proof terms,
  and a type checker.
  
  This allows to use Cyp
  for auto-grading exercises where the goal is to complete programs and proofs
  that are partially given by the instructor, as terms with holes.
  Since this allows holes in programs, type-checking becomes essential.
  Before, Cyp assumed that the program was written by
  a type-correct instructor,  and therefore omitted type-checking of proofs.
  Cyp gracefully handles incomplete student submissions.
  It accepts holes temporarily, and checks complete subtrees fully.
  
  We present basic design decisions, make some remarks on implementation,
  and include example exercises from a recent course
  that used Cyp as part of the Leipzig Autotool auto-grading system.
\end{abstract}

\section{Introduction}\label{sec:intro}

When teaching programming, it is important to underline
that each (sub)program has a specification,
and requires a proof of actually meeting it---and that the
best way to write a program is to write its proof first.

For instance, here is a programming-by-proving exercise.
Assume that Peano numbers are known:
\begin{verbatim}
data N = Z | S N

doubleN :: N -> N
doubleN Z = Z
doubleN (S x) = S (S (doubleN x))
\end{verbatim}
The topic is binary numbers now:
\begin{verbatim}
data B = Zero | Even B | Odd B
\end{verbatim}
Their semantics is given by translation to Peano numbers:
\begin{verbatim}
value :: B -> N
value Zero = Z
value (Even x) = doubleN (value x)
value (Odd x) = S (doubleN (value x))
\end{verbatim}
The goal is to implement the successor function for the binary representation:
\begin{verbatim}
succB :: B -> B
succB Zero = _
succB (Even x) = _
succB (Odd x) = _
\end{verbatim}
and to prove it correct:
\begin{verbatim}
Lemma succ : forall b :: B : value (succB b) .=. S (value b)
Proof by induction on b :: B
...
QED
\end{verbatim}
The goal here is to derive the program
from the specification (lemma \verb|succ|)
by writing the proof (replacing the dots ``\verb|...|'')
and filling holes (underscores) in the program to make the proof work.

For instance, for case \verb|b = Odd x| of the inductive proof,
we expect the following chain of reasoning,
starting with the instantiated right-hand side of the lemma.
We already use Cyp notation for a proof by rewriting:
\begin{verbatim}
                     S (value (Odd x))
(by def value)   .=. S (S (doubleN (value x)))
(by def doubleN) .=. doubleN (S (value x))
(by IH)          .=. doubleN (value (succB x)) 
(by def value)   .=. value (Even (succB x))    -- P
(by def succB)   .=. value (succB (Odd x)) 
\end{verbatim}
Up to point $P$, all steps are straightforward.
The induction hypothesis is denoted by \verb|IH|.
From point $P$ to the final expression (the left-hand side of the lemma),
we have the creative step: \verb|succB| should have equation
\begin{verbatim}
succB (Odd x) = Even (succB x)
\end{verbatim}
to make the proof work. The all too common approach
of writing the program first, and the proof later,
is characterized as ``putting the cart before the horse''
by Dijkstra~\cite{EWD:EWD1305}. Development of the proof
is the horse that pulls forward the development of the program.

The above example exercise is taken
from the ``Fortgeschrittene Programmierung'' lecture,
taught by one of the authors (Schwarz)
in Sommersemester 2020.
Student solutions for this exercise are automatically checked
with our extension~\cite{Renz20,cyp-htwk}
of the Cyp system~\cite{cyp-tum,cyp-ethz},
integrated into Leipzig Autotool~\cite{autotool}.

In the present paper, we briefly present
our existing auto-grader for programming exercises (Section~\ref{sec:autotool}),
then summarize Cyp (Section~\ref{sec:cyp}).
We then explain the design of our extensions: holes (Section~\ref{sec:holes})
and types (Section~\ref{sec:types}).
These extensions were developed in the master's thesis~\cite{Renz20}
of one of the authors (Renz).
Finally we report on our experience of using Cyp (Section~\ref{sec:exp}).
Some implementation detail is discussed in the Appendix.

\section{Preliminaries: Auto-Grading of Programming Exercises}\label{sec:autotool}

Autotool~\cite{autotool} is an auto-grader that started out in 1999
as a CGI script written by one of the authors (Waldmann)
to assist students executing derivations in Hilbert calculus.
Since then it has grown into a system supporting different exercise types.
It consists of web front-ends for students and instructors,
a semantics back-end that does the actual grading,
and a persistence component.
Autotool's primary functionality is:

\begin{itemize}
\item The instructor configures and publishes exercises.
  The students submit solutions.
  Autotool checks solutions semantically---as opposed to just checking
  syntactically whether some known good solution was met.
  This allows for detailed error messages
  that still do not give away the correct solution.
\item Exercises can be solved during a certain time period (set by the instructor).
  For students to pass the exercise, they have to submit a correct solution
  at least once in that period. Number of submissions is unlimited,
  and experimentation is encouraged, as students will practice
  reading error messages.
\item Students can compete for and check on an (pseudonymized) leaderboard,
  ordered by size of submission, or some other problem-specific parameter.
  This motivates the search for better or more elegant solutions.
\item For some exercise types, Autotool generates
  individual problem instances for each student. This does not apply
  to the programming and proving exercises described here.
\end{itemize}

We are already using Autotool
for auto-grading of Haskell programming exercises~\cite{WaldmannWFLP17}.
An instance of an exercise is a program text with holes for terms.
The student has to fill in the holes.
The text contains an expression \verb|test :: Bool|
(without holes, of course).
Autotool will check that the submission is type-correct,
and that \verb|test| evaluates to \verb|True|.
The implementation uses the Glasgow Haskell Compiler\cite{ghc}'s API
via the Mueval library~\cite{mueval}.
For notating and evaluating tests,
we use the Leancheck library~\cite{leancheck}.

The introductory example can be framed as a program-and-test exercise
if we replace the lemma with 
\begin{verbatim}
test :: Bool
test = holds 1000 $ \ (x :: B) -> value (succB x) == S (value x)
\end{verbatim}
Note the syntactic and semantic similarity.
``lambda'' takes the role of \verb|forall|.
This is exactly the point of property-based testing~\cite{DBLP:conf/icfp/ClaessenH00}.

For actually running the exercise, we add these declarations
\begin{verbatim}
{-# language DeriveGeneric, PatternSignatures #-}
import Test.LeanCheck
import Test.LeanCheck.Generic
data N = .. deriving (Eq)
data B = .. deriving (Show, Generic)
instance Listable B where tiers = genericTiers
\end{verbatim}
We need
\verb|Eq| for \verb|N| to evaluate the condition in the test,
\verb|Generic| for \verb|B| to generate test data,
and \verb|Show| for \verb|B| to print arguments for failed tests.

Type-directed generation of test cases
is an impressive showcase for the power of typeclasses in Haskell.
We make a point of including the complete source code in the exercise.
That way students can run this on their own machine,
and can also copy the approach to other code.
This would not be possible with a hidden implementation of testing.
We abhor hidden extra test cases even more.

In the above code,
it is %
sad that the natural declaration \verb|(x :: B)|
can only be written after enabling language extension \verb|PatternSignatures|,
which is deprecated, and will be replaced by \verb|ScopedTypeVariables|.
This cannot be comprehended. Where is the type variable?

\section{Preliminaries: Cyp - Check Your Proof}\label{sec:cyp}

For student exercises (and elsewhere),
testing is good, and widely used. Proving is better, but rarely done.
There seems to be a  disconnect between proving
(student writes homework proof on blackboard) and ``actual programming''
(student solves programming exercise that is auto-graded).

Cyp is a proof checker
that verifies proofs about Haskell-like programs.
It was developed by Dominik Durner and Lars Noschinski at TU Munich~\cite{cyp-tum},
and later by Dmitriy Traytel at ETH Zurich ~\cite{cyp-ethz}.
We now present the basics of Cyp by example,
and show our extensions in later sections.

Cyp was first mentioned 2014 in a report~\cite{DBLP:conf/haskell/BlanchetteHNNT14}
on teaching programming with Haskell to a large audience:
``We developed tool support for simple proofs by induction,
in the form of a lightweight proof assistant
and will integrate it into our homework submission system''.

Why not use a full proof checker, like Agda, Coq, Isabelle, PVS?
There is no publication on Cyp by the original authors,
so we have to guess a little.
A full proof checker may appear to require too much investment
of instructors' and students' time---to learn techniques and technicalities
that are perceived as ``not programming''.
Agda is perhaps closest to Haskell, but was not available widely at the time.
Also, a full proof checker may be hard to integrate in an auto-grader.

Isabelle had been used for auto-grading,
e.g., with Praktomat~\cite{praktomat16}
but that required heavy scaffolding and sandboxing
and even then, did not allow for full integration:
``we currently do not check what the students actually proved,
for that we still manually check their submission,
but this way any sorries or other problems are detected.''~\cite{Breitner16}

Cyp's design was influenced by
the Isar structured proof language~\cite{DBLP:conf/tphol/WenzelP06}
for Isabelle~\cite{DBLP:journals/fac/Nipkow89}.
The main features of Cyp are:
\begin{itemize}
\item a program is a collection of 
  \begin{itemize}
  \item 
  algebraic data types (ADT) (Haskell: \verb|data| declarations), and
\item 
  functions defined by term rewriting system
  (Haskell: oriented equations)
  \end{itemize}
\item a specification (lemma or axiom) is an equation,
\item a proof may use term rewriting,
  extensionality for functions,
  case distinction (for the constructors of an ADT),
  and induction (following the ADT).
\end{itemize}
We give examples for Cyp's proof language.
These are sometimes incomplete but the semantics should be obvious.
Full example texts and syntax and semantics specification
are available in the source repository.
In the following, all typing judgments (\verb|:: Bool|, etc.)
are only available (and then, required) in Cyp-Leipzig.
We discuss typing in Section~\ref{sec:types}.

\paragraph{Proof by Rewriting.}
A sequence of steps, where each step connects two terms
by applying a rule that comes from a function definition,
or from a lemma.
\begin{verbatim}
Lemma succ_eq_plus_one: forall a :: N: S a .=. plus (S Z) a
Proof by rewriting
                    S a
  (by def plus) .=. S (plus Z a)
  (by def plus) .=. plus (S Z) a
QED
\end{verbatim}
Cyp is not fully explicit here:
the student just names the function (here, \verb|plus|).
Cyp will determine the actual rule to use (here, \verb|plus| has two rules),
and the location and direction of its application,
via pattern-matching all rules and subterms.
We guess that authors wanted to avoid extra syntax for locations.
We consider introducing syntax for naming a rule
and specifying direction.

\paragraph{Proof by Extensionality.}

Add a symbolic argument to prove equality of functions.
\begin{verbatim}
Lemma: not . not .=. id
Proof by extensionality with x :: Bool
  Show: (not . not) x .=. id x
  ...
\end{verbatim}

\paragraph{Proof by Case Analysis.}

A complete case analysis for the constructor in the root of the term.
Continuing the previous example,
\begin{verbatim}
Proof by case analysis on x :: Bool
  Case True
    Assume  AS: x .=. True
    Then Proof ...
\end{verbatim}

\paragraph{Proof by Induction.}

Structural induction on terms.
The induction hypothesis has to be written explicitly.

\begin{verbatim}
Lemma symdiff_sym: forall x :: N, y :: N: symdiff x y .=. symdiff y x
Proof by induction on x :: N generalizing y :: N
  Case Z
    For fixed y :: N
    Show: symdiff Z y .=. symdiff y Z
    Proof ... QED
 Case S x
    Fix x :: N
    Assume IH: forall y :: N: symdiff x y .=. symdiff y x
    Then for fixed y :: N
    Show: symdiff (S x) y .=. symdiff y (S x)
    Proof ... QED
QED
\end{verbatim}
Notation \verb|Proof ... QED|
is valid in our extension, see Section~\ref{sec:holes}.

\paragraph{Restrictions.}

Cyp is not a full proof checker,
but a tool to teach equational logic and structural induction.
With that goal,
it keeps both programming and proving as simple as possible,
and we do not intend to change that.
\begin{itemize}
\item Definitions are global.
  There are no \verb|let|, or \verb|where| bindings,
  and no \verb|case| or \verb|lambda| expressions.
  This allows for a straightforward application of rewrite rules.
\item Propositions (called lemmas) are equations,
  with universal quantification over their variables.
  Cyp has no propositional logic or predicate logic.
  We did introduce mandatory type declarations for variables in lemmas,
  see Section~\ref{sec:types},
  making explicit the universal quantification.
\item All proof steps have to be written down. This is in contrast to
  interactive proof assistants that fill in proof steps,
  e.g., as Isabelle does with ``\verb|by auto|''.
\end{itemize}
The next two restrictions could possibly be lifted,
as we discuss in Section~\ref{sec:conc}.
\begin{itemize}  
\item Cyp programs, like Haskell programs, have no guarantee of termination.
  Still, equational reasoning for such programs
  is ``morally correct''~\cite{DBLP:conf/popl/DanielssonHJG06}.
\item The semantics of pattern matching in Cyp programs is different
  from the operational semantics of Haskell.
  If a function $f$ is defined by several equations,
  Haskell semantics is to try them one after another,
  until a matching left-hand side is found.
  In a Cyp proof by rewriting, \verb|(by def f)| is correct
  if any of the defining equations of $f$ can be used.
\end{itemize}

The following properties of Cyp allow for an easy integration
into systems for automated grading of homework:
\begin{itemize}
\item 
Proof checking is pure, in the Haskell sense: it does not involve IO.
Running Cyp in an auto-grader cannot change the system state
(on the disk, in the data base), and cannot leak secrets,
so it does not require sandboxing.
\item 
Proof checking is strict, in the sense of ``strong'':
any attempt to cheat, by leaving holes in the submission,
will be detected, see Section~\ref{sec:holes}.
\end{itemize}

\section{Holes}\label{sec:holes}

We mentioned earlier (Section~\ref{sec:autotool})
that we already have auto-grading for ``blueprint'' programming exercises
where students replace holes by expressions.
For auto-grading Cyp exercises, we copy that idea,
and we also allow holes in proofs.
We describe the resulting syntax and semantics.

\paragraph{Design Goals for Holes.}
The syntactic function of a hole 
is that it marks the place in a problem instance
where the student has to fill in something.
See \ref{subsec:match} for details on syntactic matching.

Holes do have semantics as well:
Cyp should not reject them, but accept temporarily,
by assuming the hole can be filled,
and continue checking other parts of the proof.
This allows for incremental development.

If holes were rejected outright,
then the student would first replace
each such hole with some expression (or proof)
that is syntactically valid,
but (most probably) semantically wrong.
By accepting the hole, we relieve the student of such rote edits.

We have a similar design for our Haskell exercises:
the (expression) hole is actually \verb|undefined|,
which GHC accepts and type-checks.
(We should switch to underscore, which recent GHCs
accept with \verb|-fdefer-type-errors|.)

\paragraph{Holes for Expressions, Names of Rules, and Rewrite Steps.}
An underscore (\verb|_|) may appear instead of a sub-expression:
in the right-hand side of a function definition,
or in an equational proof.
An underscore can also be used in \verb|(by _)|
as a placeholder for the name of a lemma, axiom, or definition
to be applied in a rewrite step.
In a proof by rewriting, three dots (\verb|...|) denote any chain of rewrite steps
where neither rules nor expressions are specified.

Expression holes have type \verb|forall a. a|
(as does \verb|undefined| in Haskell) when typechecking.
When checking the validity of a chain of equations,
Cyp will assume that anything can be rewritten into a hole,
and out of a hole.

Any rewrite step $t_1$ \verb|(by _)| $t_2$ will be accepted.
Cyp will not check that there is indeed a rule that can be
substituted for the hole.

\begin{example}\label{ex:holes:rewrite}
  This is similar to one problem we used in a lecture, see Section~\ref{sec:exp}.
\begin{verbatim}
e :: T
axiom assoc: forall x :: T, y :: T, z :: T :
  times x (times y z) .=. times (times x y) z
axiom neutral_right: forall x :: T : times x e .=. x
axiom neutral_left: forall x :: T : times e x .=. x
axiom square:  forall x :: T : times x x .=. e
Lemma  : forall x :: T, y :: T : times x y .=. times y x
Proof by rewriting
                   times x y
  (by _ )      .=. times (times x y) e
  (by square)  .=. _
  (by assoc)   .=. _
  (by assoc)   .=. _
  (by assoc)   .=. times (times x (times (times y y) x)) (times y x)
  ...             .=. times y x
QED
\end{verbatim}
We expect the student to infer \verb|(by neutral_right)|
for the first rewrite step, as no other axiom matches.
Then, \verb|(by square)| can only match
if \verb|e| is rewritten to some square,
which must be guessed to be \verb|times (times y x) (times y x)|,
and a strong hint is given by the expression
that has to be reached via associativity.
From looking at \verb|(times y y)|,
the student should guess that \verb|(by square)| comes next,
and continue with the greedy approach of always simplifying when possible.
The exercise is solved by applying 
\verb|neutral_left|, \verb|square|, \verb|neutral_left|.
By formulating the exercise with the given hints,
we have removed the ``group theoretic'' aspects
and reduced it to a pattern matching drill, on purpose.
\end{example}

\paragraph{Holes for Proofs.}

In proofs,
three dots (\verb|...|) denote any number of missing lemmas,
missing proofs (also sub-proofs) or missing cases
in proofs by case analysis, and proofs by induction.
In practice (see Section~\ref{sec:exp}),
we did not use exercises with missing lemmas.
For proofs by case analysis and by induction,
we did use heavily the option of giving one case nearly completely,
and leaving the other(s) for the student to fill in.

\begin{example}\label{ex:holes:proof}
  The case for constructor \verb|S| is missing in the following:
\begin{verbatim}
data N = Z | S N

plus :: N -> N -> N
plus Z y     = y
plus (S x) y = S ( plus x y )

Lemma plus_Z: forall x :: N : plus x Z .=. x
Proof by induction on x :: N
  Case Z
      Show: plus Z Z .=. Z
      Proof ... QED      
  ...
QED   
\end{verbatim}
\end{example}

When Cyp encounters a missing proof 
it is treated as if it were correct, much like Isabelle treats \verb|sorry|.
When a proof by case analysis or by induction has a hole for missing cases,
Cyp will assume that the present cases are complete.

\section{Types}\label{sec:types}

Originally, Cyp did not have types for programs or for proofs,
except for type judgments to determine what constructors
need to be handled in proofs by case analysis and by induction.
These judgments were checked only implicitly:
if one constructor was used, all others from the same \verb|data| declaration
had to be handled as well.

Contrary to what the student should expect from Haskell,
there was no explicit type checking,
and the notation was confusing:
the type defined by \verb+data List a = Nil | Cons a (List a)+
had to appear in the proof without the type argument:
\verb|Proof by induction on List xs|.
This looks like medieval (i.e., pre-1.4) Java
and the corresponding \verb|xs :: List| is a kind-error in Haskell.

\paragraph{Types for programs.}

Cyp did not type-check the program
since it was assumed to be given by the instructor, 
who could ensure correct typing by extra means
(divination, or GHC perhaps),
and the student could not change the program.

Our version of Cyp allows holes in the program
that the student can fill with arbitrary expressions.
For sanity, and for consistency with Haskell, we type-check programs.
We use the ``Typing Haskell in Haskell'' library~\cite{thih},
restricted to Cyp's Haskell subset (e.g., without type classes).
An expression hole is treated as \verb|_ :: forall a . a|.
This allows to accept programs with holes.

\paragraph{Types for lemmas.}

Cyp did not type-check proofs
since it was assumed that nothing bad could happen
when applying rules (from a program that was known to be type-correct)
to terms from goals (also type-correct).

\begin{example} This assumption, however, is wrong, as shown by the following:
\begin{verbatim}
data U = U

Lemma eek: x .=. y
Proof by case analysis on x :: U
  Case U
    Assume AS_x: x .=. U
    Then Proof by case analysis on y :: U
      Case U
        Assume AS_y: y .=. U
        Then Proof by rewriting
                       x
         (by AS_x) .=. U
         (by AS_y) .=. y
QED QED QED
\end{verbatim}
The problem is that Lemma \verb|eek|
is implicitly assumed to be polymorphic in $x$ and $y$,
while it was only proved for monomorphic $x$ and $y$ of type \verb|U|.
We can then prove \verb|False .=. True| by applying \verb|eek|.
\end{example}

We solve this problem by requiring every lemma to be explicit
about the types of its variables,
and then we type-check its proof in that context.
The problematic proof
\begin{verbatim}
data U = U

Lemma eek: forall x :: a, y :: a : x .=. y
Proof by case analysis on x :: U
...
QED
\end{verbatim}
will now get rejected, since the type (\verb|U|) of \verb|x| in the discriminant
is different from the (polymorphic) type of \verb|x| in the lemma.

\paragraph{Local names in lemmas.}
Explicit types also solve the following problem:
is \verb|e| a variable or a constant in:
\begin{verbatim}
axiom right_neutral : times x e .=. x
\end{verbatim}
In original Cyp, this \verb|e| is a local variable,
unless there is some globally defined \verb|e|,
then the name will refer to that, as in:
\begin{verbatim}
e = e  ;  axiom right_neutral : times x e .=. x
\end{verbatim}
This is inconsistent with local declarations
shadowing global declarations in Haskell.
We now write
\begin{verbatim}
axiom right_neutral : forall x :: T : times x e .=. x
\end{verbatim}
Since \verb|e| is not declared to be local in the axiom,
it is a reference to a global name.

\paragraph{Types in equational proofs.}

We check that all terms in an equational proof have identical type,
under the typing assumptions for the free variables of the lemma.

This allows to give precise and early error messages,
where otherwise the error would likely have manifested
in a more obscure way, e.g., by there not being a fitting rewrite rule.

\paragraph{Typed application of rules.}

When a rule from a lemma or from an axiom
is applied in a proof by rewriting,
Cyp determines, by syntactic matching,
what subterm $t_1$ is rewritten to what $t_2$,
and under what substitution of the lemma's variables.
We use our type-checking mechanism to see
if the variables' and the terms' types unify.

\begin{example}
Assume we proved \verb|eek| for the correct type:
\begin{verbatim}
data U = U
Lemma eek: forall x :: U, y :: U: x .=. y 
Proof ... QED
\end{verbatim}
Then the following attempt to apply \verb|eek|
\begin{verbatim}
data Bool = False | True
Lemma contradiction: False .=. True
Proof by rewriting
               False
  (by eek) .=. True
QED
\end{verbatim}
will get rejected,
because we obtain the substitution
that maps \verb|x| to \verb|False|
and \verb|y| to \verb|True|
for rewriting the left- to the right-hand-side
using \verb|eek|,
and the types \verb|x :: U| and \verb|False :: Bool|
as well as the types of \verb|y :: U| and \verb|True :: Bool|
do not unify.
\end{example}

\section{Experience Report}\label{sec:exp}

Cyp, with our extensions, was being used in the 
``Fortgeschrittene Programmierung'' (Advanced Programming) lecture
for Computer Science B. Sc. students in their 4th semester.
(The recent addition of typed lemmas was not available.)
The lecture follows~\cite{WaldmannWFLP17}
and was taught by one of the authors (Schwarz) in 2020.
While auto-grading for Haskell exercises was used before,
this was the first instance of using Cyp as well.
Because of Corona lock-down, the lecture was virtual.
This made auto-grading of part of homework all the more important.
(There were other homework and discussions via internet platforms, 
shared documents, and audio conferencing.)

\paragraph{How Cyp was used.}
We installed GHC (with Leancheck) and Cyp in the computer pool
(on Debian GNU/Linux machines) and students logged in via SSH.
Students were also encouraged to install GHC and compile Cyp on their home computers.
The instructor used the pool installation in tutorial presentations
on BBB (Big Blue Button) conferences, on a server hosted by our department: 
share the view of one terminal window,
SSH to a computer in the pool,
start \verb|tmux| and split the window,
run an editor in one half, and show GHCi or Cyp output in the other half.

While we have integrated Cyp in Autotool,
we always maintain a command line version of Cyp with identical semantics.
It can be called with one argument (a module to be checked)
or two (a blueprint module, and a solution module).

The recommended ``IDE for Cyp'' is the command
\begin{verbatim}
while (inotifywait -q *.[bc]yp); do cyp -m ex.byp ex.cyp; done
\end{verbatim}
where
\verb|ex.byp| (\underline{b}lueprint for c\underline{yp}) is the problem instance,
with the holes, downloaded from Autotool;
and \verb|ex.cyp| is the student's working file.
Each time this file is saved from inside the editor,
Cyp will check it (locally). Finally, the student uploads the solution to Autotool,
where it is checked again, and the outcome is stored.

\paragraph{Types of exercise used.}

Over 14 weeks of teaching, we posed Cyp exercises on
\begin{itemize}
\item proof by rewriting (only), cf. Example~\ref{ex:holes:rewrite}
\item proof by case analysis (and rewriting), e.g.,
  \begin{example}
\begin{verbatim}
data Bool = False | True  

xor :: Bool -> Bool -> Bool
xor False False = False  ; xor False True  = True
xor True  False = True   ; xor True  True  = False

Lemma xor_sym : forall x :: Bool, y :: Bool : xor x y .=. xor y x
Proof by case analysis on x :: Bool 
 Case False
   Assume  AX: x .=. False
   Then Proof by case analysis on y :: Bool
   Case False
     Assume  AY: y .=. False
     Then Proof by rewriting
                  xor x y
      (by AX) .=. _
      (by AY) .=. _
      (by AY) .=. _
      (by AX) .=. _
     QED
   Case True
     Assume  AY: y .=. True
     Then Proof by rewriting
              xor x y
      ... .=. xor y x
     QED
  QED
  ...
QED
\end{verbatim}
  \end{example}
\item proofs by induction on \dots
  \begin{itemize}
  \item 
  Peano numbers, cf. Example~\ref{ex:holes:proof},
  and another on associativity and commutativity of addition,
  with implementation of addition given, and assuming
\begin{verbatim}
axiom plus_Z: plus x Z .=. x
axiom plus_S: plus x ( S y ) .=. S ( plus x y )
\end{verbatim}
\item singly linked lists:
\begin{verbatim}
len ( append xs ys ) .=. plus ( len xs ) ( len ys )
\end{verbatim}
\item binary trees, prove \verb|mirror (mirror t) .=. t|
 \item a representation of binary numbers, see Section~\ref{sec:intro}.
  \end{itemize}
\end{itemize}

\paragraph{Statistical data on auto-graded exercises.}
In total, 39 mandatory problems were assigned in Autotool, 
twelve of them contained proofs in Cyp and
15 were Haskell programs with holes.

Two problems %
were given with complete solutions beforehand.
They were also used by the instructor during an online session
to demonstrate the usage of the Cyp installation in our computer pools.

Most of the problems require just one proof, some up to three.
In problems that contain more than one proof,
the first proof was sometimes given with a near-solution and the others
without further hints. %

Two problems required
proofs by rewriting,
three by case analysis,
and seven by induction on
Peano numbers,
binary numbers,
lists,
or trees.

Near-solutions containing only \verb|_| holes for expressions were provided for
three proofs.
The other problems contained \verb|...| holes for whole proofs.
In around half of them, the proof method (rewriting, case analysis,
induction) was required.

Some problems were given in two versions. %
On the one hand, the students have to fill holes for expressions in a Haskell
program to satisfy the specification that is tested by Leancheck.
On the other hand, they have to prove in Cyp that their code satisfies
this specification.
In these cases, the solution to the first problem should not be given
in the setting of the Cyp program.
Therefore, we use \verb|_| holes in programs in the Cyp problem
in addition to the holes in proofs. 

\paragraph{Student performance  and evaluation.}

Among the 95 participants of the lecture, 71 solved at least 19
Autotool problems correctly.
Out of these 71, all solved at least two Cyp problems and most solved
at least half of all Cyp problems.

In addition to Autotool, students had to solve exercise sheets and discussed
their solutions online. Some of these problems also contained proofs.
We observed that many student solutions were given in Cyp format and had been
checked by Cyp, even if we did not explicitly require this.

During the last lecture, the instructor asked the 43 participants via
an online poll:
\emph{Did Cyp exercises in Autotool help you to understand how to write proofs?}
Out of 28 answers, 27 said yes and one said no.

Therefore, we now consider using Cyp examples and exercises
earlier in the curriculum, maybe even before a thorough introduction to the
Haskell programming language.

\section{Discussion and Conclusion}\label{sec:conc}

We discuss two topics for further development of Cyp.

\paragraph{Operational Semantics of Cyp Programs.}

An interesting connection is that
while we want to replace a property-based test by a proof of the property,
the Isabelle IDE automatically can derive and execute such a test from a lemma.

Could we do the same for Cyp, as the example from the introduction suggests?
We can certainly add \verb|deriving (Eq, Show, Generic)|
to every \verb|data| declaration,
and mechanically generate a testable property from a lemma.

But what is the relation between lemmas that are ``operationally true'' in Haskell
and lemmas that we can prove?
Cyp will accept
\begin{verbatim}
f :: Bool -> Bool ; f False = False ; f False = True
Lemma False .=. True
Proof by rewriting   False
  (by def f) .=. f False
  (by def f) .=. True
QED
\end{verbatim}
but the second proof step does not have an operational equivalent in Haskell.

We could enforce operational semantics of top-down matching.
This would complicate rule application in equational proofs,
as we have to check that earlier rules don't match.
Given this function declaration
\begin{verbatim}
f (S x) y = _
f x (S y) = _
\end{verbatim}
we cannot rewrite the expression \verb|f a (S b)| (in some equational proof)
with either rule, since \verb|a| needs to be elaborated first.
Cyp could rewrite patterns by removing cases that are covered earlier.
The second rule then comes out as \verb|f Z (S y)|.
A compiler will do this~\cite{DBLP:conf/fpca/Augustsson85}
when generating code for an abstract machine,
but Cyp generates code that must be used by students.

It seems better to avoid this problem
by enforcing that patterns of each function definition are disjoint.
Students can get the same behaviour from GHC
via \verb|-Woverlapping-patterns|.

\paragraph{Termination.}

Totality of functions is a subject of teaching.
Cyp's functions are not total.
(The previous paragraph shows that they aren't even functions.)
One reason for non-totality is incomplete patterns in function definitions,
which we could check, as GHC already does with
\verb|-Wincomplete-patterns|.
Another reason is non-termination, coming from recursion,
which is unrestricted both in Haskell and in Cyp.
\begin{itemize}
\item We could prohibit recursion altogether.
  This is a drastic measure, and then we lose the motivation for induction.
  It still can be useful, e.g., to force the student to implement
  a specification by using \verb|fold| (with given axioms),
  and not by ad-hoc recursion.
\item We could restrict recursion to those cases
  where if a well-founded monotone order of function calls
  can be determined by Cyp,
  or is provided by the student.
\end{itemize}
The Cyp-style solution would certainly be to make this
simple (no complicated order, just the subterm relation)
and explicit (annotating the decreasing argument).

\paragraph{Conclusion.}
We extended the Cyp proof checker
with a pattern matcher for program and proof terms,
and with a type checker.
This allows to auto-grade student exercises
on completing implementations and proofs of Haskell-like programs.

Cyp can be used stand-alone (on the command line),
and in our auto-grading system Autotool.
We used Cyp successfully in a lecture.

\bibliographystyle{myplainurl}

\newpage
\appendix
\section{Technicalities}\label{sec:tech}

We comment on some design decision in the implementations of Cyp.
Since there is no formal publication on the original,
we can only make educated guesses.
We still present it, as an exercise in understanding
design decisions in real-life Haskell software,
and also as basic documentation for some changes that we executed.

In very general terms, the design of a proof checker,
or any other static analyzer, is:
``parse, then analyze, then print result or error message''.
The semantics of matching, type checking, and proof checking are well understood.
The result is just the empty tuple \verb|()|,
as we only want to know whether the input was accepted.
But some details of parsing in Cyp are interesting,
and so is printing of error messages.

\subsection{Parsing}\label{subsec:parse}

Since Cyp is about proving properties of programs in (a subset of) Haskell,
the authors decided (we guess) 
to use a full Haskell parser~\cite{hses} for theories,
and a hand-written (combinatorial) parser for proofs (only).

Using the full Haskell parser avoids duplication of work,
and ensures that Cyp programs are syntactically valid Haskell.
But it also creates problems:
\begin{itemize}
\item 
  Error messages from the full Haskell parser
  sometimes refer to concepts that are (syntactically) forbidden in Cyp.
  This confuses the student.
\item These parts are not completely separated:
  (Haskell) expressions appear in equational proofs.
\item Mixing parsers is hard, since this full Haskell parser
  is not composable.
\end{itemize}
We expand the last item:

We expect  a parser to have type \verb|String -> m (result,String)|,
returning the result, and the part of the input that was not consumed.
These parsers can be composed. But the full Haskell parser is not composable
(probably because it is based on a shift/reduce model).
It wants to completely read a string that contains a valid expression
(or declaration, etc.) so we have to trim the string on the outside.

Cyp uses a simple algorithm for trimming:
each (Haskell) expression extends to end-of line,
or to \verb|.=.| in a lemma.
This  makes the implementation look awkward,
as it contains a lot of actual string processing
with jumping back and forth looking for separator symbols \verb|\n| and \verb|.=.|

We decided to rip out the full Haskell parser, and replace it
with a hand-made combinatorial parser for the ``Cyp subset of Haskell''.
This is not much extra work (the subset is small)
and it gives us uniform, composable parsers with better error messages,
and reduced compile time, as \verb|haskell-src-exts|,
the basis of  \verb|haskell-src-exts-simple|, is hard for GHC.

For backward compatibility, we still use ``expression to end-of-line'' syntax,
and we implemented combinators
\begin{verbatim}
to_eol :: Parser a -> Parser a
to_eol p = inline (p <* eof)

inline :: Parser a -> Parser a
inline p = do
  src <- getInput
  let (pre, post) = span (\c -> not $ elem c "\n\r") src
  setInput pre
  x <- p
  pre' <- getInput ; setInput $ pre' <> post ; whiteSpace
  return x
\end{verbatim}
with typical usage
\begin{verbatim}
eqnSeqFromList <$> to_eol eterm
    <*> ( many $ (,) <$> link <* reservedOp ".=." <*> to_eol eterm )
\end{verbatim}

The move away from \verb|haskell-src-exts(-simple)|
was also suggested by the fact that \verb|haskell-src-exts|
is ``on life support''. Maintenance was deemed too costly,
as the library duplicates functionality that is already in GHC.
Instead, GHC's Haskell parser is available in a library (\verb|ghc-lib-parser|).
Switching Cyp to that would also create work.

\subsection{Printing}\label{subsec:print}

Another aspect of syntactical processing is
(pretty) printing of parts of the input, in error messages.

In the naive approach, a parser produces an AST (abstract syntax tree).
Later, if some semantic analysis detects some error in some subtree,
this subtree should be printed. But if the tree is truly abstract,
we forgot how it actually looked in the input!
So the pretty-printer needs to do some work,
e.g., to introduce parentheses according
to precedence and associativity of operators.
Indeed, Cyp has all these pretty-printers (called ``unparsers'') for ASTs.

There is an alternative: we keep enough information in the AST nodes
so that we can always show them in their original form,
and never need to compute a ``pretty'' layout.
It is, in fact, confusing for the student if Cyp (or any compiler)
complains about code that is not recognizable as part of the source.

We could attach the actual input slice to each AST node.
But it is customary to keep the full input elsewhere,
and attach location (range) information of type \verb|Span| to AST nodes,
by implementing the \verb|HasSpan| type class:
\begin{verbatim}
import Text.Parsec
data Span = Span SourcePos SourcePos
class HasSpan a where
  mspan :: a -> Maybe Span
  set_span :: Maybe Span -> a -> a
\end{verbatim}
In the parser, we typically have code like
\begin{verbatim}
prop :: Parser RawProp
prop = with_span $ Prop <$> term <*> ( reservedOp ".=." *> term )
\end{verbatim}
using a function
\begin{verbatim}
with_span :: HasSpan a => Parser a -> Parser a
with_span p = do
  start <- getPosition ; res <- p ; end <- getPosition
  return $ set_span (Just $ Span start end) res
\end{verbatim}

Then the pretty-printer can exactly re-create the input.
But then another problem appears: the printer needs access to the input.
This is solved by our (re)design of the ``Error monad''
where all semantical analysis take place:
it is enhanced by a Reader monad for the input string.
This monad originally had a function
\begin{verbatim}
err :: Doc -> Err a
\end{verbatim}
to show an error message and stop processing.
We can now use a version with source spans (extracted from ASTs)
\begin{verbatim}
errs :: [Span] -> Doc -> Err a
\end{verbatim}
and the corresponding source slices will be rendered with the message.

\subsection{Hiding Annotations by Pattern Synonyms created automatically}\label{subsec:anno}

We now need to carry around the source spans
(extra syntactical information)
without destroying legacy code (for semantics):

For example, Cyp has this type (not all constructors shown)
\begin{verbatim}
data AbsTerm a
    = TermHole
    | Application (AbsTerm a) (AbsTerm a)
    | Const Name
    | ...
\end{verbatim}
With source spans, it would be
\begin{verbatim}
type AbsTerm a = AbsTermP (Maybe Span) a
data AbsTermP p a
    = TermHoleP p
    | ApplicationP p (AbsTerm p a) (AbsTerm p a)
    | ConstP p Name
    | ...
\end{verbatim}
but now all pattern matches on \verb|AbsTerm a|
would be broken by the extra argument.
Perhaps we should have written each constructor with named arguments,
but we'd have to touch all code, and the result would be questionable,
since constructor application would then look much less
like the function application that it really is.

GHC Haskell has ``pattern synonyms''~\cite{DBLP:conf/haskell/PickeringEJE16}
to solve exactly this problem: We can write
\begin{verbatim}
pattern Application :: AbsTerm a -> AbsTerm a -> AbsTerm a
pattern Application f a <- ApplicationP _ f a
  where Application = ApplicationP Nothing
\end{verbatim}
and similarly for the other constructors.
In fact, \verb|haskell-src-exts|
has an AST type like \verb|AbsTermP| (with source span),
and \verb|haskell-src-exts-simple|
provides pattern synonyms without.

We are using exactly that idea, made easier by the following tool:
for all AST types (Terms, Types, Proofs),
we derive the annotated type, and the synonyms,
from the ``plain'' definition (original \verb|AbsTerm| in  the example)
by a Template Haskell program (it creates extra source code,
in memory, during compilation, resulting in actual object code).
This creates the interesting situation that Cyp now has modules
(containing the original AST definitions)
that only need to be compiled so that they can be analyzed
when Template Haskell runs,
but that will not be part of the executable.

In all, this design allows to keep intact
all Cyp code that pattern-matches on ASTs.
Still the extra source-span information is available when we need it
(not by pattern matching, but by calling \verb|mspan|).

\subsection{Generic Traversal for Matching Holes}\label{subsec:match}

We mention another example of generic programming in Cyp:
the matching of a student's submission with an instructor's blueprint
(with the holes) is done by a generic traversal.
The basic idea for matching term $t$ against blueprint $b$ is:
if $b$ is a hole, we accept;
and if $b$ is  not a hole,
we check that the roots of $b$ and $t$ are identical,
and we match corresponding children recursively.

While original Cyp had type-specific matchers (and pretty-printers),
we use  the capabilities of \verb|Data.Data|
for run-time reflection on (names of) constructors.
We still use Haskell's static typing, so we can be sure
that $b$ and $t$ are of the same type always,
and have the same number of children when we recurse.

We need to adapt this basic idea for (proof) holes
that may be replaced by a list of things.
If we allow this in full generality, we need to decide, for example
\begin{verbatim}
[ ... , b1, ..., b2, ... ]  matches [ t1, t2, t3, t4, t5, t6 ]
\end{verbatim}
Now, $b_1$ matches $t_i$ for what $i$?
This is similar to the longest common subsequence problem,
and can be solved similarly, in polynomial time,
but it requires search. If the search fails,
what should we print as an error message?
So we are inclined to severely restrict multi-holes in lists,
allowing them to occur at most once.

Our implementation uses code originally written by Bertram Felgenhauer
for Autotool's Haskell Blueprint exercise.
It uses a monad \verb|M| for combining results from comparing subtrees.
A computation in \verb|M a| can succeed, fail, or decline.
Failure is final (it ends all processing).
Declination leads to the next alternative (of \verb|msum| in the following)
to be tried. There are two essential combinators:
\begin{itemize}
\item sequential composition (\verb|(>>=)|):
  if left-hand side succeeds, evaluate right-hand side
\item alternative composition (\verb|mplus|):
  if left-hand side declines, evaluate right-hand side.
\end{itemize}
The monad also holds a state of the pair of last locations that we have seen
(from the root). This is useful: not every node has location info,
and for them, we will use the source span of the parent node.

The main function is
\begin{verbatim}
match  :: forall a b . (Data a, Data b) => a -> b -> Match b
match x y = do
  msum
    [ processSpanSub x y
    ...
    , matchList @ParseLemma  (cast x) (cast y)
    , matchHole @RawTerm (cast x) (cast y)
    , matchHole @ParseProof (cast x) (cast y)
    ...
    , if toConstr x == toConstr y
      then void $ gzipWithM' match x y else continue
    , failLoc $ "abstract syntax sub-trees of type "
      <> (text $ dataTypeName $ dataTypeOf x)
      <> " have different roots: "
      <> (text $ show $ toConstr x) <> ", " <> (text $ show $ toConstr y)
    ]
  return y
\end{verbatim}
Here, we are using generic helper functions (\verb|matchHole|)
instantiated at certain types (\verb|@RawTerm|).

This method allows to remove type-specific traversal code.
We have not completely done so, as we are still evaluating
the effect of losing type-specific error messages.

\end{document}